\documentclass[aps,prl,twocolumn,nopacs,superscriptaddress]{revtex4}
\usepackage[utf8]{inputenc}
\usepackage[T1]{fontenc}

\usepackage{graphicx}  % needed for figures
\usepackage{dcolumn}   % needed for some tables
\usepackage{bm}        % for math
\usepackage{amssymb}   % for math
\usepackage{amsmath}
\usepackage{units}
\usepackage[english]{babel}
\usepackage[dvipsnames]{xcolor}
\usepackage{xfrac}
\usepackage{multirow}

\usepackage[%
colorlinks=true,
urlcolor=RoyalBlue,
linkcolor=RoyalBlue,
citecolor=RoyalBlue,
]{hyperref}

\usepackage{sidecap}
\sidecaptionvpos{figure}{t}  

\usepackage[type1]{libertine}                                    
\usepackage{textcomp}% Required to get special symbols
\usepackage[scaled=.85]{beramono}% Typewriter font
\usepackage[libertine,cmintegrals,cmbraces,vvarbb,slantedGreek]{newtxmath}
\usepackage[scr=boondoxo]{mathalfa}% Extra math symbols
\usepackage{bm}% Extra bold faces
\usepackage[lf]{carlito}

\usepackage{braket}
\usepackage{bm}

\makeatletter
\DeclareRobustCommand{\cev}[1]{%
  \mathpalette\do@cev{#1}%
}
\newcommand{\do@cev}[2]{%
  \fix@cev{#1}{+}%
  \reflectbox{$\m@th#1\vec{\reflectbox{$\fix@cev{#1}{-}\m@th#1#2\fix@cev{#1}{+}$}}$}%
  \fix@cev{#1}{-}%
}
\newcommand{\fix@cev}[2]{%
  \ifx#1\displaystyle
    \mkern#23mu
  \else
    \ifx#1\textstyle
      \mkern#23mu
    \else
      \ifx#1\scriptstyle
        \mkern#22mu
      \else
        \mkern#22mu
      \fi
    \fi
  \fi
}
\makeatother

\newcommand{\spinhalf}{spin-\sfrac{1}{2}}

\newcommand{\panel}[1]{(#1)}
\newcommand{\panelcaption}[1]{(#1)}
\newcommand{\panelsubcaption}[1]{(#1)}

\begin{document}

\title{Electric Control of Spin Transitions at the Atomic Scale}

\author{Piotr Kot}
\affiliation{Max-Planck-Institut f\"ur Festk\"orperforschung, Heisenbergstraße 1,
70569 Stuttgart, Germany}
\author{Maneesha Ismail}
\affiliation{Max-Planck-Institut f\"ur Festk\"orperforschung, Heisenbergstraße 1,
70569 Stuttgart, Germany}
\author{Robert Drost}
\affiliation{Max-Planck-Institut f\"ur Festk\"orperforschung, Heisenbergstraße 1,
70569 Stuttgart, Germany}
\author{Janis Siebrecht}
\affiliation{Max-Planck-Institut f\"ur Festk\"orperforschung, Heisenbergstraße 1,
70569 Stuttgart, Germany}
\author{Haonan Huang}
\affiliation{Max-Planck-Institut f\"ur Festk\"orperforschung, Heisenbergstraße 1,
70569 Stuttgart, Germany}
\author{Christian R. Ast}
\email[Corresponding author; electronic address:\ ]{c.ast@fkf.mpg.de}
\affiliation{Max-Planck-Institut f\"ur Festk\"orperforschung, Heisenbergstraße 1,
70569 Stuttgart, Germany}

\date{\today}

\begin{abstract}
    Electric control of spins has been a longstanding goal in the field of solid state physics due to the potential for increased efficiency in information processing. This efficiency can be optimized by transferring spintronics to the atomic scale. We present electric control of spin resonance transitions in single molecules by employing electron spin resonance scanning tunneling microscopy (ESR-STM). We find strong bias voltage dependent shifts in the ESR signal of about ten times its linewidth, which is due to the electric field induced displacement of the spin system in the tunnel junction. This opens up new avenues for ultrafast control of coupled spin systems, even towards atomic scale quantum computing, and expands on understanding and optimizing spin electric coupling in bulk materials. 
\end{abstract}

\maketitle

\section{Introduction}

Spintronics and the concept to control spin and magnetic properties using electric fields have been on the forefront of solid state research for the past several decades with the promise to increase efficiency in data processing \cite{matsukura_control_2015,kane_silicon-based_1998,atzori_second_2019,eerenstein_multiferroic_2006}. Different concepts have been considered such as the spin transistor \cite{datta_electronic_1990,gregg_spin_2002,hirohata_review_2020,sugahara_spin-transistor_2010}, the spin Hall effect \cite{kato_observation_2004,sinova_spin_2015}, dopants in silicon \cite{laucht_electrically_2015,tosi_silicon_2017,asaad_coherent_2020}, and magnetic molecules \cite{trif_spin-electric_2008,liu_quantum_2021,liu_electric_2019,fittipaldi_electric_2019,robert_polyanisotropic_2019,palii_electric_2014,cardona-serra_electrically_2015,gaita-arino_molecular_2019,godfrin_operating_2017}. Specifically, spin-electric control allows for superior scalability and switching as electric fields are more easily contained and faster to manipulate than magnetic fields. This type of processing could be further optimized by transfering it to the atomic scale, for which scanning tunneling microscopy (STM) is an ideal platform in realizing such a goal. Specifically, the combination of electron spin resonance spectroscopy (ESR) with STM has expanded the sensitivity of ESR to atomic scale spin systems, and has enhanced the attainable energy resolution of STM well into the neV range \cite{baumann_electron_2015,paul_generation_2016,natterer_upgrade_2019,weerdenburg_scanning_2021,drost_combining_2022}.

As the manipulation capabilities in STM are mostly based on electrical control, implementing sizeable atomic scale electrical spin control can become not only possible with ESR-STM, but also quite effective. The applied bias voltage typically induces a very strong electric field between the tip and sample due to the extremely small gap of only a few \AA ngströms \cite{girard_electric_1993}. Moreover, ESR spectra are typically acquired by sweeping the microwave frequency or the magnetic field, so that the bias voltage essentially becomes a free parameter to be tuned. However, so far the bias voltage in ESR-STM has not been employed for spin manipulation. 

\begin{figure}
\includegraphics[width=1\linewidth]{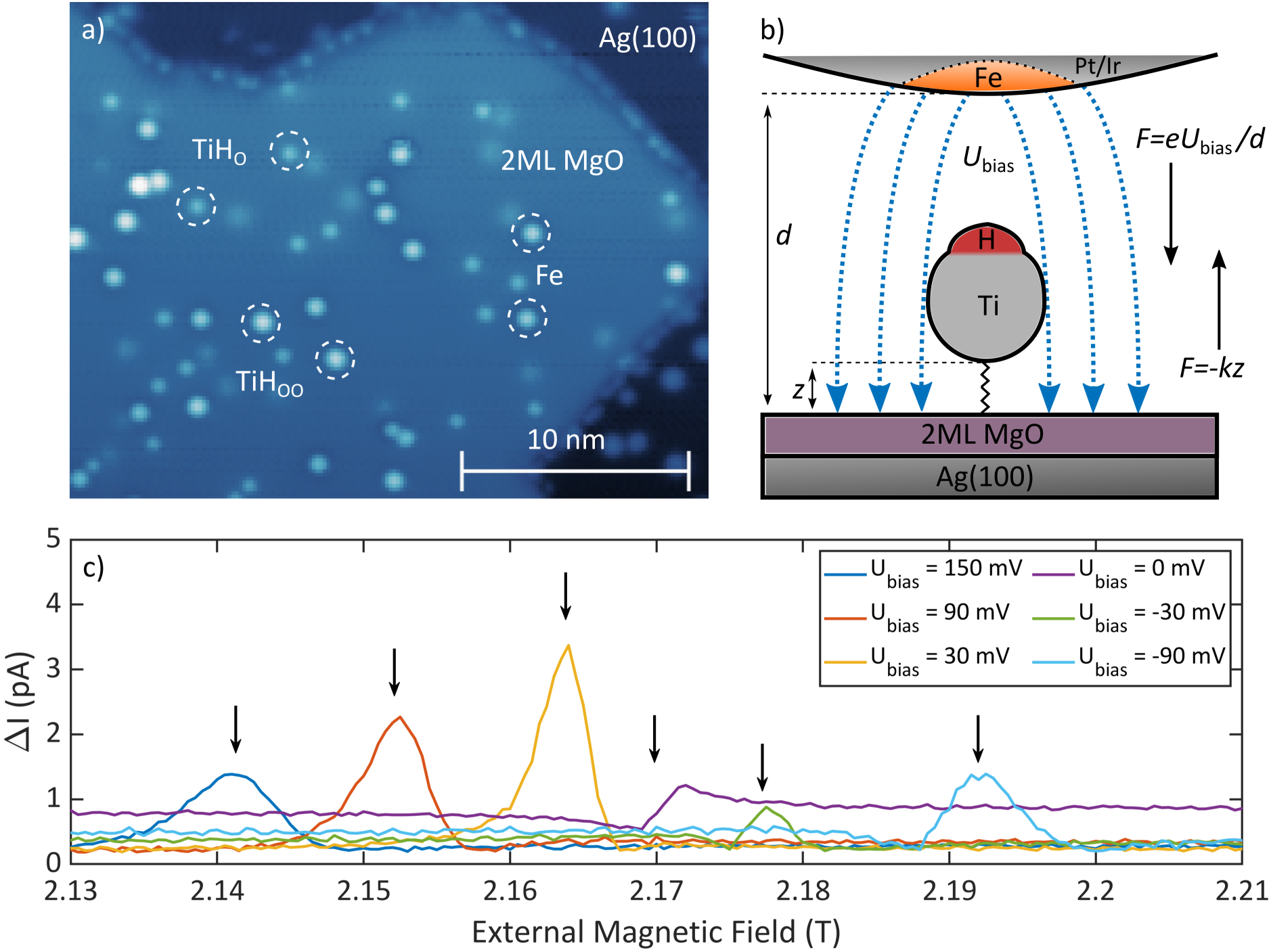}
    \centering
    \caption{\textbf{ESR on TiH molecules} \panelcaption{a} STM topography of 2\,ML MgO on a Ag(100) substrate decorated with individual TiH molecules and Fe atoms ($U_{sp} =$ 100\,mV, $I_{sp} =$ 20\,pA). The different species are labelled and circled accordingly. \panelcaption{b} Schematic of the tunnel junction during the ESR experiment. Force vectors representing the electric force induced by the bias voltage and elastic force of the Ti-MgO bond are shown. Additionally, the electric forces may act on the Ti-H bond.  \panelcaption{c} Magnetic field sweeps performed at different STM junction bias voltages $U_\text{bias}$ ($U_\text{sp} = 100$\,mV, $I_\text{sp} = 250$\,pA, $f_\text{rf} = 61.545$\,GHz, $U_\text{rf}$ = 20\,mV). The ESR peak positions are labelled with black arrows.}
    \label{fig:Fig1}
\end{figure}

In this study, we exploit the bias voltage as an electrical means for direct manipulation of spin transitions. We use a TiH molecule on an insulating MgO layer (see Fig.\ \ref{fig:Fig1}\panel{a}) to demonstrate a direct tuning of the $g$-factor and the tip magnetic field. In this system, the resonance peak shifts by many line widths within a bias voltage range of $240\,$mV (see Fig.\ \ref{fig:Fig1}\panel{b}), which is much stronger than what has been predicted for this system (on a different adsorption site) \cite{ferron_single_2019} or previously measured in bulk systems \cite{liu_quantum_2021}. We explain this effect by the strong electric field in the tunnel junction induced by the applied bias voltage and felt by the dipolar TiH molecule. A change in the electric force shifts the equilibrium position of the TiH molecule, resulting in the $g$-factor being modified and the molecule feeling a different magnetic field from the spin-polarized tip. The $g$-factor is, in part, modified due to a change in the crystal field felt by the TiH \cite{ferron_single_2019}.

\section{Voltage Dependent ESR-STM}

\begin{figure}
\includegraphics[width=1\linewidth]{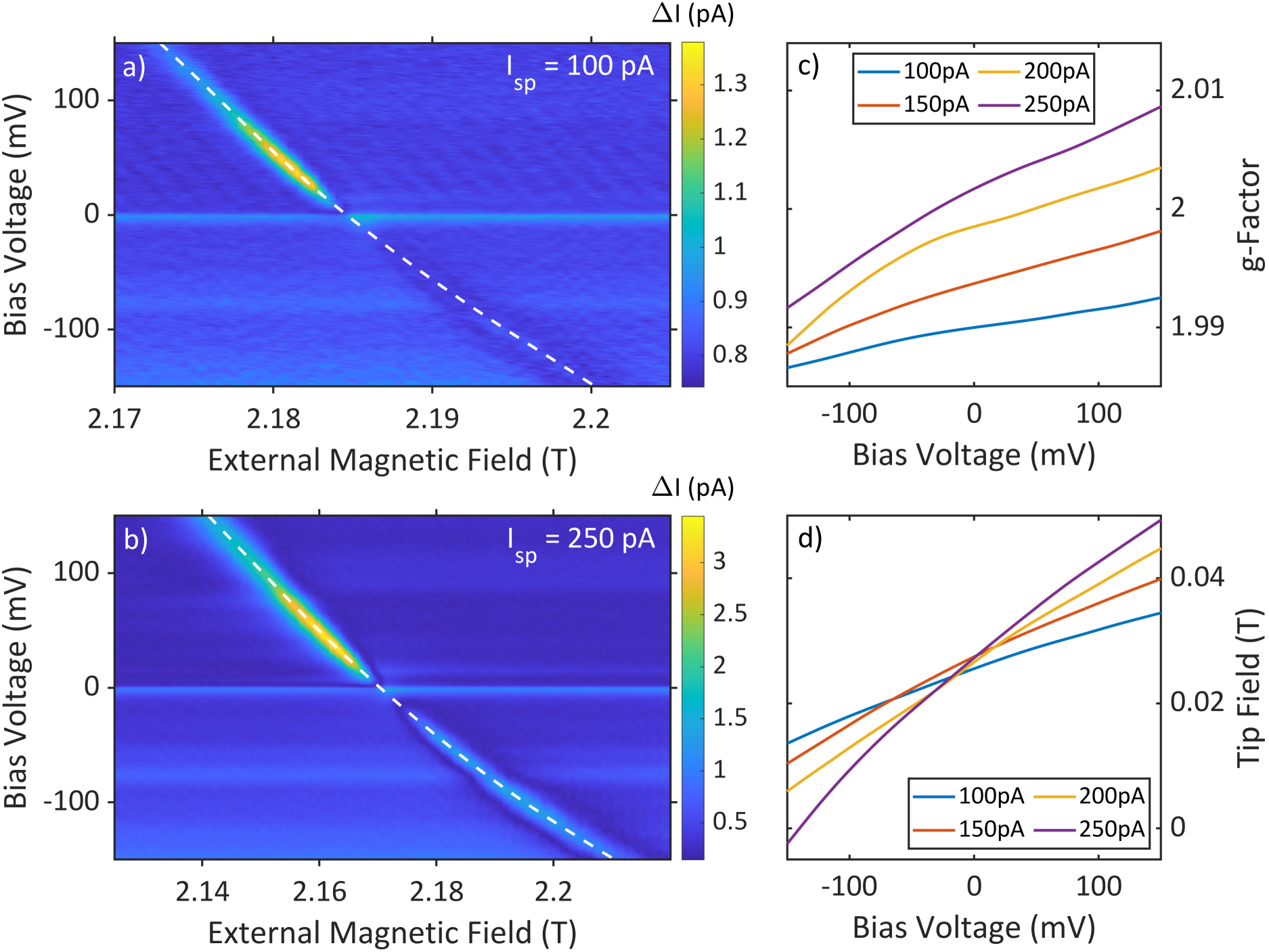}
    \centering
    \caption{\textbf{Voltage dependence of the ESR signal} \panelcaption{a}-\panelcaption{b} Magnetic field/bias voltage sweeps performed at two different current set points ($U_\text{sp} = 100$\,mV, $f_\text{rf} = 61.545$\,GHz, $U_\text{rf} = 20$\,mV, (a) $I_\text{sp} = 100$\,pA, (b) $I_\text{sp} = 250$\,pA). White dashed lines show a spline fit to the ESR peak positions as a function of bias voltage. \panelcaption{c}-\panelcaption{d} Extracted $g$-factor and tip field vs.\ bias voltage at four current set points.}
    \label{fig:Fig2}
\end{figure}

The measurements were done on TiH molecules that adsorb on the bridge-site between two O atoms of the MgO double layer. They are labelled as TiH\textsubscript{OO} in Fig.\ \ref{fig:Fig1}\panel{a}. Varying the bias voltage continuously, we observe the evolution of the ESR peak as a function of both bias voltage and external magnetic field at a constant microwave radiation frequency of 61.545\,GHz and a microwave amplitude of 20\,mV. This is shown for two different setpoint currents of $I_\text{sp}=100\,$pA and $I_\text{sp}=250\,$pA in Fig.\ \ref{fig:Fig2}\panel{a} and \panel{b}, respectively. Unless otherwise noted, the corresponding setpoint voltage is $U_\text{sp}=100\,$mV. The horizontal features in Fig.\ \ref{fig:Fig2}\panel{a} and \panel{b} are due to the interaction of the microwaves with the background density of states and are not related to the ESR signal \cite{tien_multiphoton_1963,kot_microwave-assisted_2020,roychowdhury_microwave_2015,peters_resonant_2020}. Comparing the slope of the ESR peak in Fig.\ \ref{fig:Fig2}\panel{a} and \panel{b}, we directly see that the change in the resonance condition is more pronounced for the higher setpoint current, which points towards an influence of the electric field rather than the bias voltage. We have obtained similar results for TiH molecules adsorbed on top of an O atom of the MgO layer (labeled TiH\textsubscript{O} in Fig.\ \ref{fig:Fig1}\panel{a}), which are presented in the Supplementary Information \cite{sinf}. 

For a more quantitative analysis of the evolution of the ESR peak, we exploit the linear dependence of the ESR resonance on the magnetic field as
\begin{equation}
    E_\text{Z} = hf_\text{res} = g\mu_\text{B}(B_\text{ext}+B_\text{tip}),
\end{equation}
where $E_\text{Z}$ is the Zeeman energy, $f_\text{res}$ is the resonance frequency, $g$ is the $g$-factor, and $B_\text{ext,tip}$ are the external magnetic field and the field of the tip felt by the spin system (henceforth the tip field), respectively. Furthermore, we assume the spin to be $S=\sfrac{1}{2}$ \cite{yang_engineering_2017},  $h$ is Planck's constant, and $\mu_\text{B}$ is the Bohr magneton. Both the tip field $B_\text{tip}$ and the $g$-factor will be a function of the applied bias voltage. Analyzing the data at different frequencies, we extract the $g$-factor and the tip field $B_\text{tip}$ dependency on the bias voltage at four different setpoint currents, which is shown in Fig.\ \ref{fig:Fig2}\panel{c} and \panel{d} (for details on the curve fitting, see the Supplementary Information \cite{sinf}). We can clearly see that both the $g$-factor and the tip field $B_\text{tip}$ monotonically increase with increasing bias voltage. This indicates that both quantities are sensitive to the changing electric field. In addition, the change is stronger at a larger setpoint current, which is consistent with our interpretation as a smaller tip-sample distance will lead to a stronger adjustment of the electric field with respect to bias voltage. 

One notable difference in the behavior of the $g$-factor and the tip field $B_\text{tip}$ is around zero bias voltage, where the effects of the electric field vanish.. Interestingly, near zero bias voltage the tip field is relatively stagnant as a function of the set point current, while the increase in the $g$-factor is comparable to non-zero bias voltages. Calculations in the literature show that the $g$-factor increases as the molecule-substrate distance decreases for TiH\textsubscript{O} \cite{steinbrecher_quantifying_2021,ferron_single_2019} (we expect a similar behavior for TiH\textsubscript{OO}). We have measured approach curves demonstrating that the molecule-substrate coupling increases as the tip-sample distance is reduced. This indicates a decrease in the molecule-substrate distance, which provides an overall consistent behavior for the increasing $g$-factor for larger set point currents (see Supplementary Information for details \cite{sinf}). Our findings show that adjusting the tip-sample distance results in changes to both the tip field and the $g$-factor. The changes due to the tip-sample distance have previously been attributed to the tip field \cite{yang_engineering_2017,seifert_single-atom_2020,seifert_accurate_2021}, while theoretical considerations of an electric field dependence have not taken changes in the tip field into account \cite{ferron_single_2019}. However, as we show here, the two effects cannot be easily separated.

To compare our results with literature, we calculate an effective frequency shift as a function of applied bias voltage of 0.83\,GHz/V and 4.3\,GHz/V for the $g$-factor and the tip field, respectively, at a setpoint current of 250\,pA. These values are orders of magnitude larger than what has recently been reported for the ESR peak shift of 5.7\,kHz/V in a bulk matrix of HoW$_{10}$ nanomagnets \cite{liu_quantum_2021}. We can reach these values because the electric field becomes extremely large between the tip and sample. Comparing the spin-electric coupling (SEC) constants, which relate the frequency shift to the applied electric field, the situation looks a bit different. For the HoW$_{10}$ nanomagnets \cite{liu_quantum_2021}, a value of 11.4\,Hz/(V/m) was reported, while we estimate values of 0.4\,Hz/(V/m) and 2.2\,Hz/(V/m) for the $g$-factor and the tip field $B_\text{tip}$, respectively, assuming a tip-sample distance of about 5\,\AA. While this indicates a more efficient coupling mechanism for the HoW$_{10}$ nanomagnets, the particular TiH system was not optimized \textit{a priori} for high SEC, so we anticipate spin systems with superior SEC to be identified in the future.

Furthermore, the response of the Zeeman splitting to an electric field has been previously calculated specifically for the TiH molecule on MgO, albeit on an oxygen site TiH rather than on bridge site TiH \cite{ferron_single_2019}. The calculated frequency shift is estimated to be about 0.2\,GHz/V, which is smaller than what we have observed experimentally. Neglecting the effect of the tip field, which was not considered in the calculations, we find a four times stronger change in the frequency shift for the $g$-factor in the experiment. We surmise that additional changes other than the crystal field gradient and the equilibrium position of the whole TiH molecule, such as a change in the Ti-H bond or simply the different adsorption site, contribute to this difference. The sensitivity of the TiH molecule to the local environment is already illustrated by changing the spin state from \sfrac{3}{2} in the gas phase to \sfrac{1}{2} upon adsorption on the surface, as well as changing the $g$-factor from about 2 to 0.6 by moving to a different binding site on the MgO  \cite{steinbrecher_quantifying_2021,burrows_spectroscopic_2005}. The ability now to tune the $g$-factor and the tip field $B_\text{tip}$ by means of the bias voltage opens up an entirely new degree of freedom for \textit{in situ} electrical manipulation of the spin transitions.

\begin{figure}
    \includegraphics[width=1\linewidth]{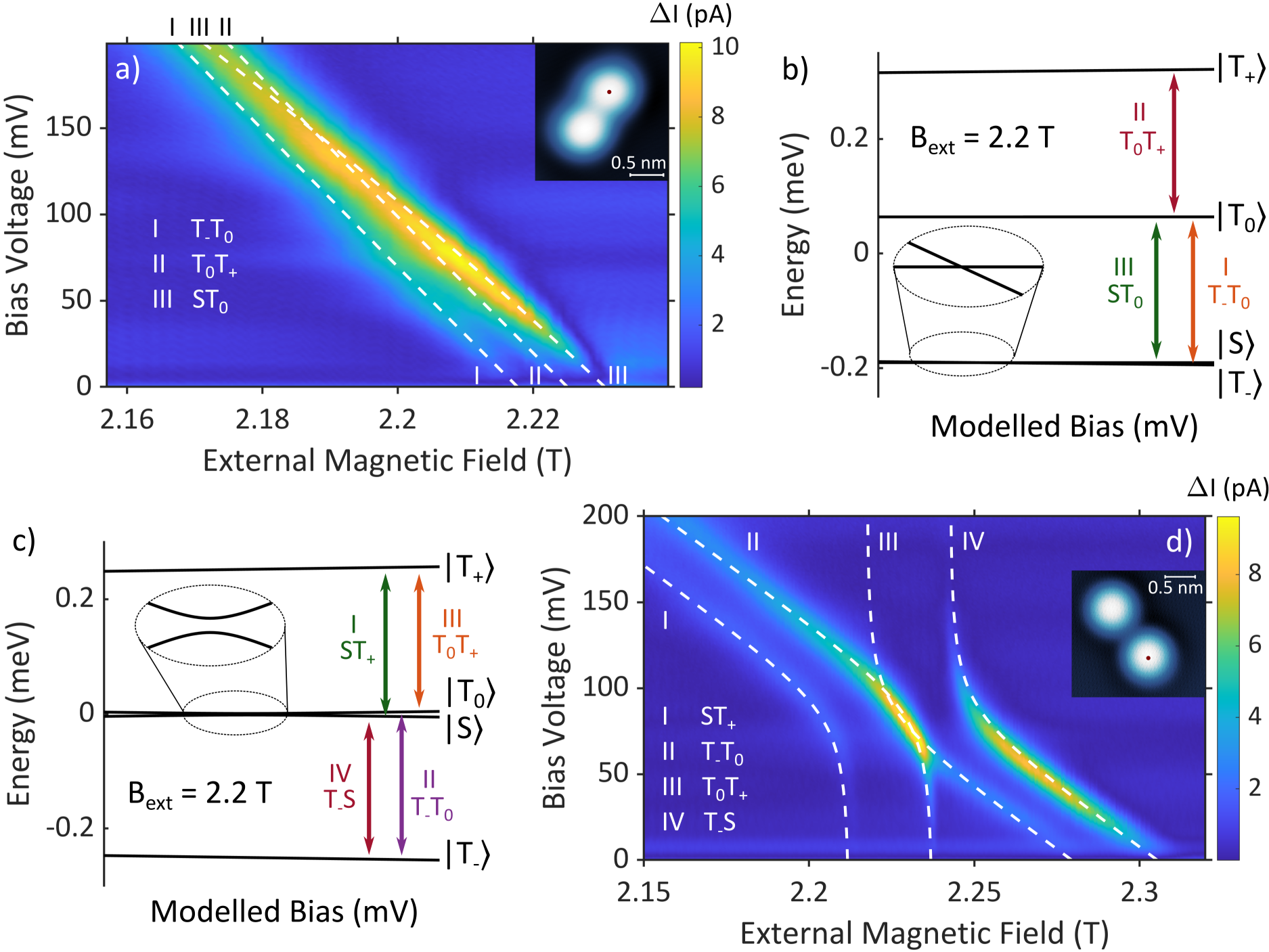}
    \centering
    \caption{\textbf{Interaction tuning in dimers} \panelcaption{a} Magnetic field/bias voltage sweep on a strongly coupled dimer ($U_\text{sp} = 150$\,mV, $I_\text{sp} = 1$\,nA, $f_\text{rf} = 61.545$\,GHz, $U_\text{rf} = 20$\,mV). The inset shows the topography of the dimer with the red dot indicating the position of the tip during measurements ($U_\text{sp} = 100$\,mV, $I_\text{sp} =$ 20\,pA). The white dashed lines are fits to the ESR peak positions corresponding to the transitions in panel \panelsubcaption{b}. \panelcaption{b} Modelled behaviour of the spin states at a constant external magnetic field that result in the ESR transitions measured by the experiment in \panelsubcaption{a}. The colored arrows show the observed transitions. \panelcaption{c} Modelled behaviour of the spin states at a constant external magnetic field that result in the ESR transitions measured by the experiment in \panelsubcaption{d}. The colored arrows show the observed transitions. \panelcaption{d} Magnetic field/bias voltage sweep showing the avoided crossing of two coupled TiH molecules ($U_\text{sp} = 100$\,mV, $I_\text{sp} = 400$\,pA, $f_\text{rf} = 61.545$\,GHz, $U_\text{rf} = 20$\,mV). The inset shows the topography of the dimer with the red dot indicating the position of the tip during measurements ($U_\text{sp} = 100$\,mV, $I_\text{sp} = 20$\,pA). The white dashed lines are fits to the ESR peak positions corresponding to the transitions in panel \panelsubcaption{c}.}
    \label{fig:Fig3}
\end{figure}

\section{Electric Control of Multiple Spins}

We demonstrate direct manipulation through SEC on two different types of dimers with different distances between the TiH molecules \cite{bae_enhanced_2018,yang_engineering_2017}. In the first example, the two bridge site TiH molecules (TiH\textsubscript{OO}) are 644\,pm apart (see inset in Fig.\ \ref{fig:Fig3}\panel{a}), such that the coupling is relatively strong ($J \approx 61.1$\,GHz). We identify three transitions in this dimer in Fig.\ \ref{fig:Fig3}\panel{a}. These transitions (labeled I, II, and III) are well separated near zero bias voltage and subsequently broaden as well as intersect as we increase the bias voltage \cite{bae_enhanced_2018}. The white dashed lines are fits to a dimer spin Hamiltonian assuming a linear dependence of the $g$-factors and the tip field $B_\text{tip}$ on the bias voltage (for details see the Supplementary Information \cite{sinf}). The corresponding energy levels at a constant external field of 2.2\,T are plotted in Fig.\ \ref{fig:Fig3}\panel{b} with the transitions being indicated. We identify transition III as a clock transition that would not be visible if the two $g$-factors in the dimer were equal \cite{bae_enhanced_2018}. Therefore, we know that the two $g$ factors are not equal even at zero bias voltage. Furthermore, as shown in Fig.\ \ref{fig:Fig3}\panel{a} we can tune transitions II and III such that they are located at the same external magnetic field value, which demonstrates that we can manipulate the spin transitions in a dimer by means of SEC.

If the two TiH molecules are 1.04\,nm apart (see inset in Fig.\ \ref{fig:Fig3}\panel{d}), the interaction between them is reduced ($J \approx 0.67$\,GHz), which shifts the energy of the singlet state $|S\rangle$ close to the triplet state $|T_0\rangle$ as shown in Fig.\ \ref{fig:Fig3}\panel{c} \cite{bae_enhanced_2018,veldman_free_2021}. The singlet state $|S\rangle$ and the triplet state $|T_0\rangle$ undergo an avoided crossing (see inset in Fig.\ \ref{fig:Fig3}\panel{c}), which can be observed experimentally \cite{veldman_free_2021}. We have tuned the tip-sample distance such that we can observe this avoided crossing in a bias voltage range between 0\,mV and 200\,mV as shown in Fig.\ \ref{fig:Fig3}\panel{d}. The four transitions that are visible in Fig.\ \ref{fig:Fig3}\panel{d} are labelled I through IV corresponding to the transitions indicated in Fig.\ \ref{fig:Fig3}\panel{c}. We can clearly see how the two pairs of transitions associated with each TiH molecule in the dimer approach the avoided crossing and separate again. The white dashed lines are fits to the same dimer spin Hamiltonian as before, just with a weaker exchange interaction, which corroborates the experimental observations (for details on the parameters see the Supplementary Information \cite{sinf}). For smaller magnetic fields below the avoided crossing, the transitions I and II are strongly influenced by the SEC, which indicates that the wave functions of the corresponding energy levels are located on the TiH molecule under the tip. As transitions III and IV are much less influenced by the applied bias voltage, we conclude that the corresponding wave functions are located on the TiH molecule next to the tip. The slope is not vertical, so we expect some influence of the electric field on the TiH molecule next to the tip about 1\,nm away. For higher magnetic fields above the avoided crossing, the situation is reversed, such that the wave functions for transitions III and IV are in the TiH below the tip and the wave functions for transitions I and II are in the TiH next to the tip.

\section{Optimizing Coherence in Coupled Spin States}

The ability to manipulate spin interactions in dimers through SEC clearly demonstrates the versatility of voltage dependent ESR-STM. However, the tunneling current itself is the biggest source of decoherence in the ESR excitation \cite{willke_probing_2018}. As a final proof-of-principle, we exploit both the bias voltage and the tip-sample distance as two degrees of freedom to move the avoided crossing to zero bias voltage, where the tunneling current is minimized and correspondingly the coherence time is maximized. This should enhance and maximize the coherent evolution of entangled states in a TiH dimer that has recently been demonstrated \cite{veldman_free_2021}.

\begin{figure}
\includegraphics[width=1\linewidth]{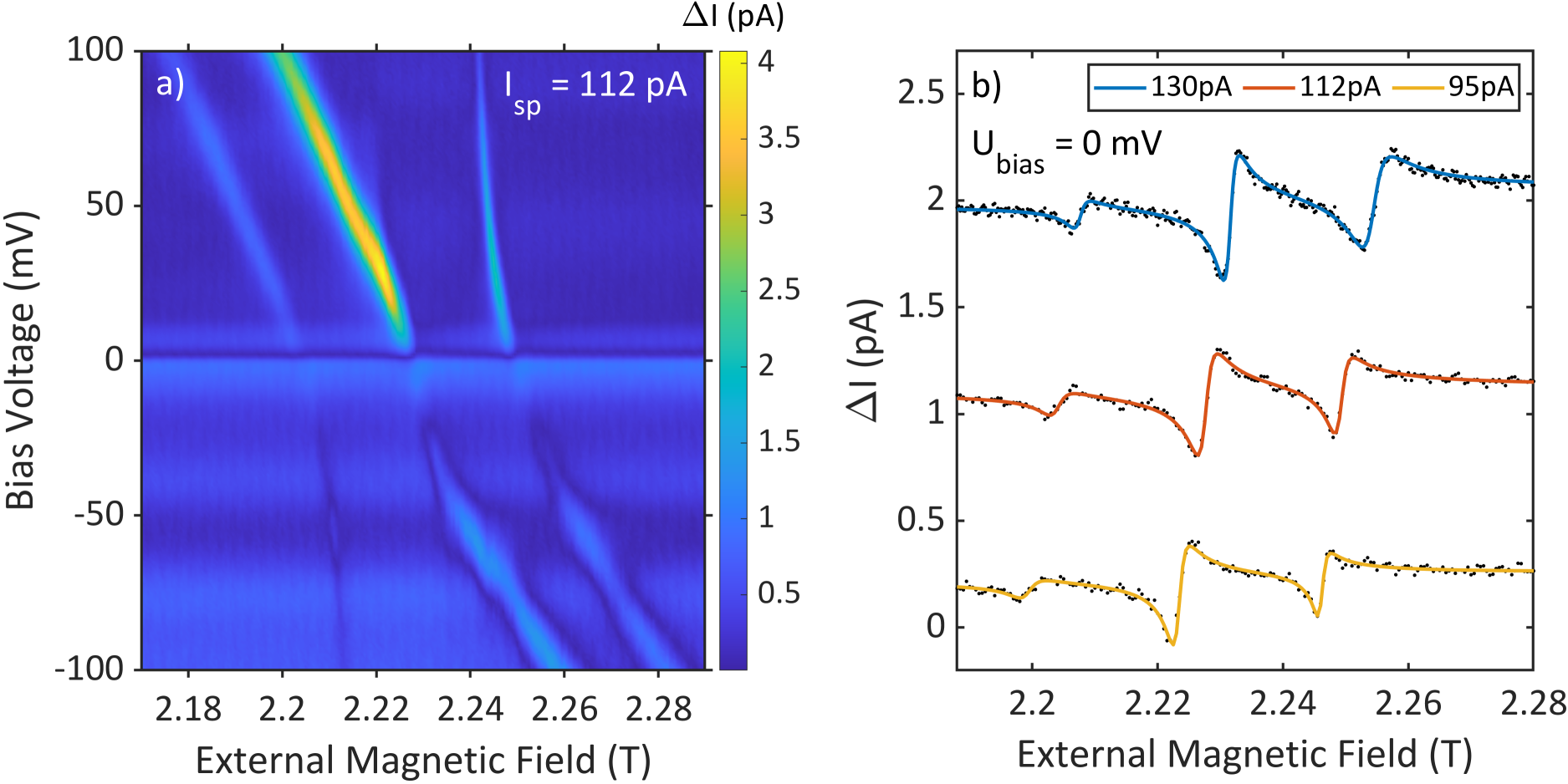}
    \centering
    \caption{\textbf{Tuning the avoided crossing} \panelcaption{a} Magnetic field/bias voltage sweep showing the avoided crossing of the TiH dimer in Fig.\ \ref{fig:Fig3}\panel{c} near zero bias voltage ($U_\text{sp} = 50$\,mV, $I_\text{sp} = 112$\,pA, $f_\text{rf} = 61.545$\,GHz, $U_\text{rf} = 20$\,mV). (b) ESR sweeps measured at zero bias showing how the resonances shift with respect to current set point ($U_\text{sp} = 50$\,mV, $f_\text{rf} = 61.545$\,GHz, $U_\text{rf} = 20$\,mV). This shows that the tip-sample distance can be adjusted to bring the avoided crossing exactly to zero bias.}
    \label{fig:Fig4}
\end{figure}

In order to move the avoided crossing of the second TiH dimer in Fig.\ \ref{fig:Fig3}\panel{d} to zero bias voltage, we increase the tip-sample distance such that the setpoint reduces from $U_\text{sp} = 100$\,mV and $I_\text{sp} = 400$\,pA to $U_\text{sp} = 50$\,mV and $I_\text{sp} = 112$\,pA. Here, the avoided crossing shifts in bias voltage when adjusting the tip-sample distance, but essentially remains at the same position in external magnetic field. Fig. \ref{fig:Fig4}\panel{a} shows the corresponding measurement, where the avoided crossing is now moved close to zero bias voltage. At zero bias voltage only the homodyne detection scheme allows to observe the ESR peaks, which typically appear as asymmetric peaks \cite{bae_enhanced_2018}. This can be seen in Fig.\ \ref{fig:Fig4}\panel{b} for three different current setpoints, where the avoided crossing is above zero voltage (blue), near zero voltage (red), and below zero voltage (yellow). The shifts of the resonances corresponding to the movement of the avoided crossing in bias voltage is clearly visible. This demonstrates that by considering the bias voltage in ESR-STM, we can manipulate spin structures in a more complex manner than previously possible.

\section{Conclusions}

The ability to tune spin transitions at the nanoscale by means of an electric field opens up many new and interesting possibilities in the atomic manipulation capabilities of STM far beyond the proof-of-principle presented here. It adds the otherwise unconsidered bias voltage to the degrees of freedom for customizing spin systems to specific needs. In this regard, the tip-sample distance, which has previously been used, and the bias voltage present ideal tuning parameters for manipulating complex spin structures. Furthermore, the bias voltage opens avenues towards a more complete understanding of the ESR mechanism in the STM and its dynamics as well as its sources of decoherence and dissipation. This becomes particularly interesting for future applications in time-resolved experiments as it enables fast switching schemes for the bias voltage, which is not possible for magnetic fields or the tip-sample distance (e.g.\ coherent evolution \cite{veldman_free_2021}, qubit operations \cite{petta_coherent_2005,he_two-qubit_2019}). Looking on a broader perspective, we have established SEC in ESR-STM, which connects to the well established field of spintronics on an atomic scale. Moreover, studying the influence of the electric field within ESR-STM opens new possibilities and a better understanding for optimizing SEC in bulk materials.

\section{Acknowledgements}

The authors would like to thank Juan Carlos Cuevas, Andreas Heinrich, Klaus Kern, Jose Lado, Sander Otte, and Aparajita Singha for fruitful discussions. We are grateful to the European Research Council (ERC) for their financial support. This study was funded in part by the ERC Consolidator Grant AbsoluteSpin (Grant No. 681164) and by the Center for Integrated Quantum Science and Technology (IQ$^\textrm{\small ST}$).

\clearpage
\newpage

\onecolumngrid
\begin{center}
\textbf{\large Supplementary Material for \\ Electric Control of Spin Transitions at the Atomic Scale}
\end{center}
\vspace{1cm}
\twocolumngrid

\setcounter{figure}{0}
\setcounter{table}{0}
\setcounter{equation}{0}
\renewcommand{\thefigure}{S\arabic{figure}}
\renewcommand{\thetable}{S\Roman{table}}
\renewcommand{\theequation}{S\arabic{equation}}

\vspace{0.5cm}

\section{Tip and Sample Preparation}

We cleaned Ag(100) in UHV by repeated cycles of Ar\textsuperscript{+} ion sputtering at 5\,kV and annealing at 820\,K. MgO was grown on the clean Ag by simultaneous evaporation of Mg onto the sample surface, leaking of O\textsubscript{2} into the UHV space, and heating of the Ag substrate. Deposition times varied from 15 to 20 minutes with the Mg Knudsen cell being heated to 500\,K, the O\textsubscript{2} being leaked to $10^{-6}$\,mbar and heating of the Ag to 520\,K. After the MgO growth, we deposited Fe and Ti on the surface using e-beam evaporators by applying an emission voltage of 850\,V and an emission current of 8.5\,mA for Fe and 19\,mA for Ti. Furthermore, the sample was kept below 16\,K during Fe and Ti deposition to ensure that the atomic species did not form clusters on the surface. The Ti species naturally hydrate due to the residual hydrogen gas found in the UHV space \cite{si_yang_engineering_2017}. To create ESR sensitive tips we picked up between one and ten Fe atoms \cite{si_baumann2015electron}. Dimers studied in this letter were either found naturally occurring on the sample or were created via atom manipulation \cite{si_yang_probing_2021}.

\section{Magnetic Field/Bias Voltage Sweeps}

We performed magnetic field/bias voltage sweeps on TiH molecules found on islands of MgO with a height of two monolayers (ML). Measurements were done by irradiating the junction at one frequency, and taking bias voltage dependent sweeps as a function of magnetic field. To minimize artifacts due to drift, we waited at least for two hours after approaching the tip and applying the microwave radiation prior to starting a sweep. To ensure that we do not drift off the molecular species under investigation, we performed atom tracking between bias sweeps while the magnetic field was ramping to the next value. In addition, we set the ramp rate of the magnet to relatively low values ($\approx 2.5$\,mT/s), ensuring minimal heating and slow adjustment of the STM junction. During each bias sweep, atom tracking was turned off and the tip position was set to hold. Lastly, we modulated the radiation at a chopping frequency of 107\,Hz and set the demodulation frequency of our lock-in amplifier to the same frequency. This way we can pick up the ESR signal of the system in the lock-in amplifier and increase our signal to noise ratio \cite{si_seifert2020single}. These sweeps took anywhere from four to twelve hours depending on the number of points being measured.

\section{Extracting ${g}$-Factors and Tip Fields}

\begin{figure}[b]
\centerline{\includegraphics[width=\columnwidth]{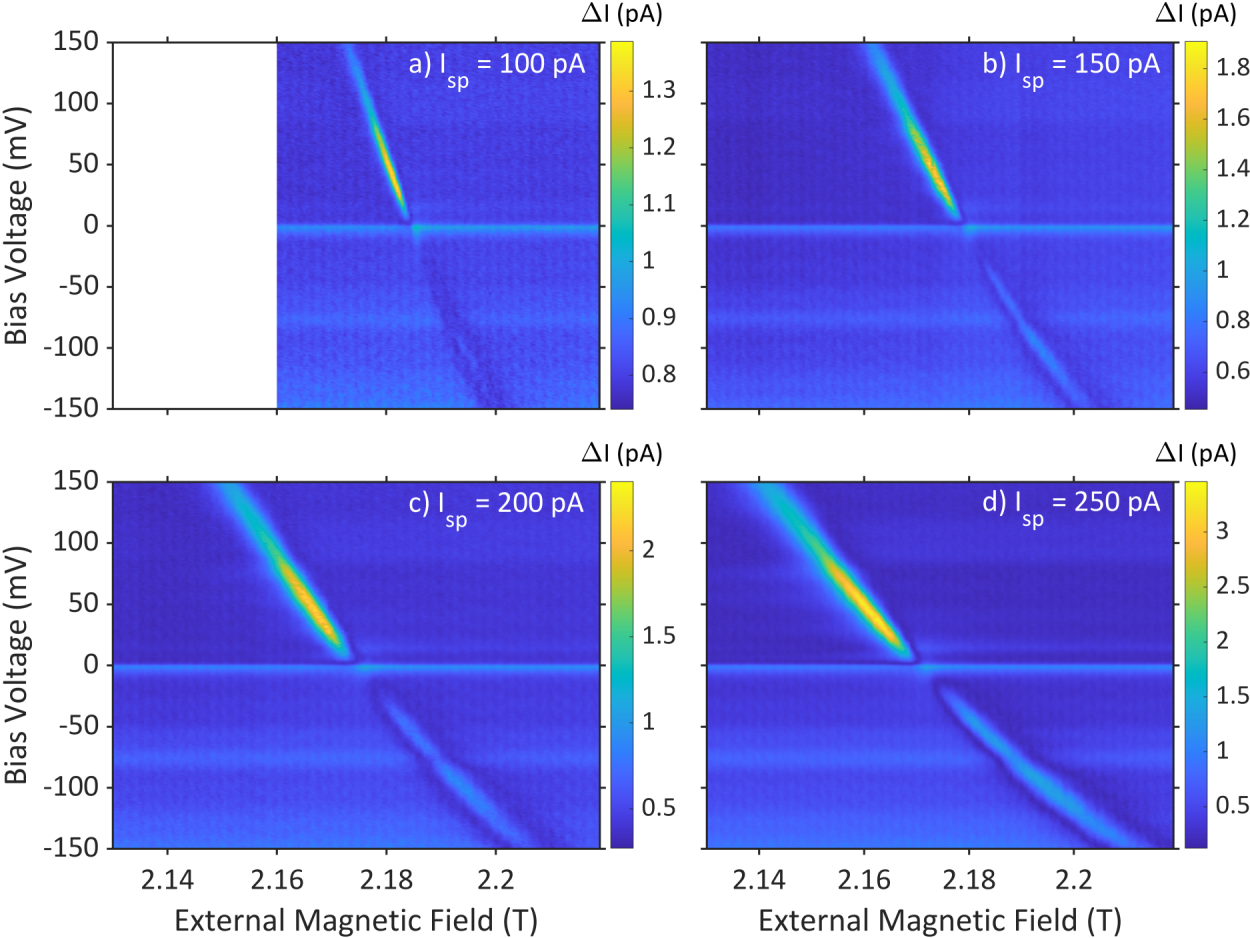}}
\caption{Magnetic field/bias voltage sweeps measured at different tip-sample distances ($U_\text{sp} = 100$\,mV, $f_\text{rf} = 61.545$\,GHz, $U_\text{rf} = 20$\,mV, \panelcaption{a} $I_\text{sp} = 100$\,pA, \panelcaption{b} $I_\text{sp} = 150$\,pA, \panelcaption{c} $I_\text{sp} = 200$\,pA, (d) $I_\text{sp} = 250$\,pA)}
\label{fig:SI5}
\end{figure}

\begin{figure}
\centerline{\includegraphics[width=\columnwidth]{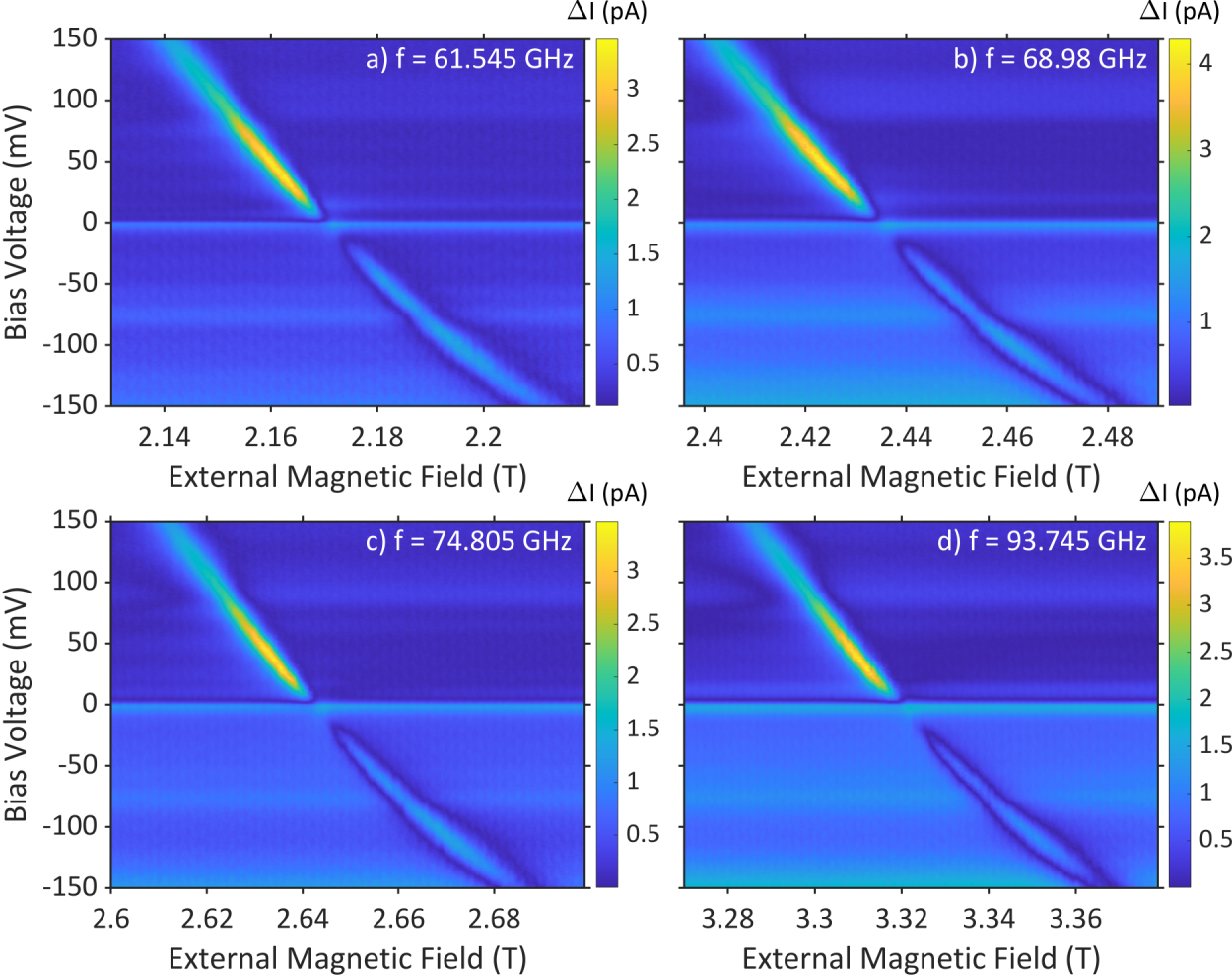}}
\caption{Magnetic field/bias voltage sweeps measured at different microwave frequencies ($U_\text{sp} = 100$\,mV, $I_\text{sp} = 250$\,pA, $U_\text{rf} = 20$\,mV, \panelcaption{a} $f_\text{rf} = 61.545$\,GHz, \panelcaption{b} $f_\text{rf} = 68.98$\,GHz, \panelcaption{c} $f_\text{rf} = 74.805$\,GHz, (d) $f_\text{rf} = 93.745$\,GHz)}
\label{fig:SI4}
\end{figure}

To extract the bias voltage dependency of the $g$-factor and the magnetic field of the tip presented in the main text in Fig.\ 2, we measured magnetic field/bias voltage sweeps on a bridge site TiH molecule (TiH\textsubscript{OO}) at four different microwave frequencies (i.e.\ Zeeman energies) and four different current set points. Fig.\ \ref{fig:SI5} shows magnetic field/bias voltage sweeps at four different current set points. We keep the $x$-axis scaling the same in all panels to more clearly show the effect of the tip-sample distance on the bias voltage shift of the ESR signal. Already there is a clear indication that the spin-electric coupling (SEC) is stronger at smaller tip-sample distances. Fig.\ \ref{fig:SI4} shows magnetic field/bias voltage sweeps measured at four different microwave frequencies. The horizontal features in all panels of Figs.\ \ref{fig:SI4} and \ref{fig:SI3} are due to the interaction of the microwaves with the background density of states and not related to the ESR signal (cf.\ \cite{si_tien_multiphoton_1963,si_kot_microwave-assisted_2020}).  

\begin{figure}[h]
\centerline{\includegraphics[width=\columnwidth]{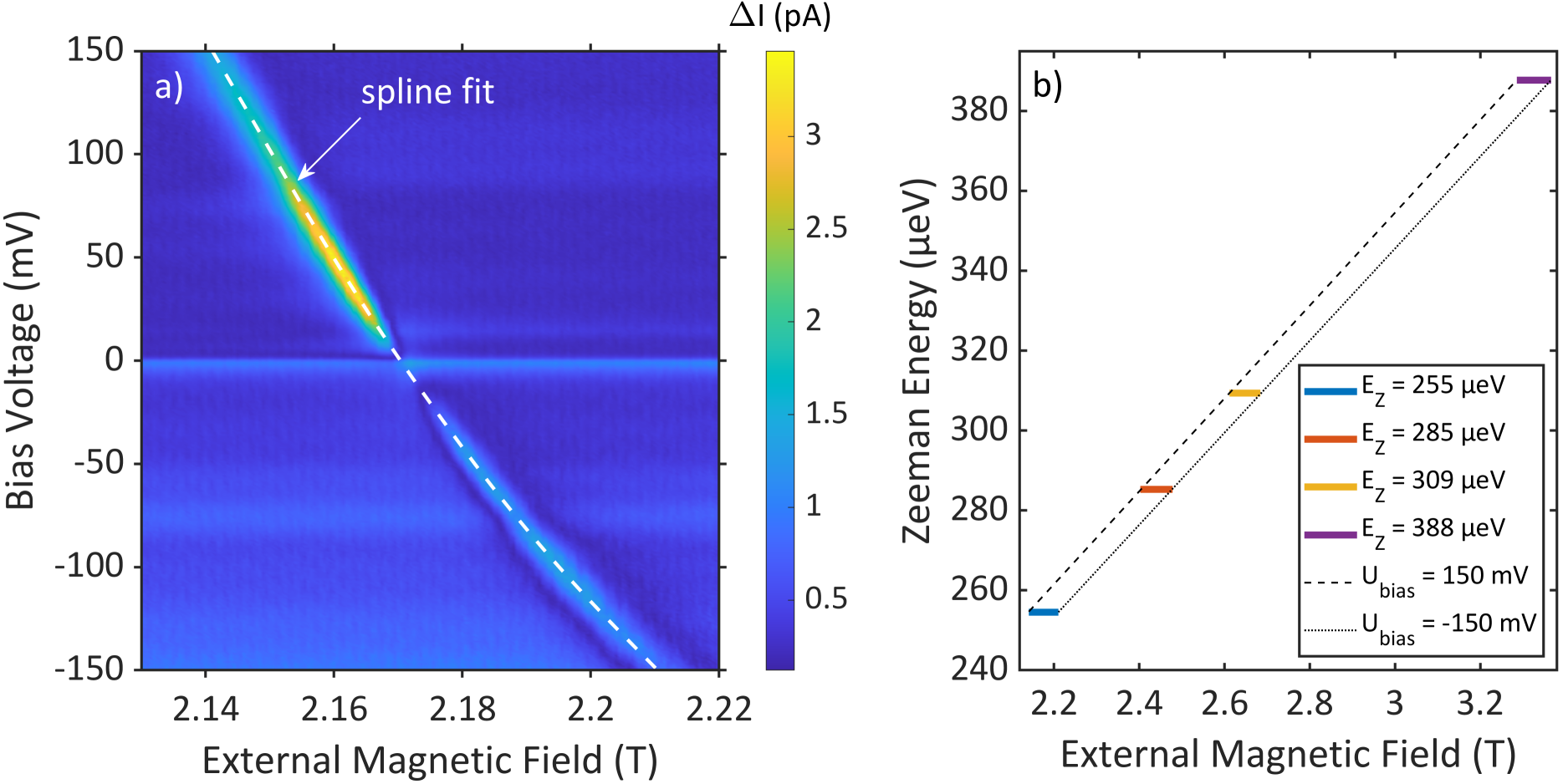}}
\caption{\panelcaption{a} Magnetic field/bias voltage sweep with a dashed line indicating the spline fit performed over the full bias voltage range. \panelcaption{b} Representation of the linear fits at each bias voltage to extract the bias voltage dependencies of the $g$-factor and tip field $B_\text{tip}$. Two linear fits are shown at 150\,mV and $-150$\,mV represented by the dashed and dotted lines, respectively.}
\label{fig:SI1}
\end{figure}

We can extract the dependencies of the g-factor and the tip field on the bias voltage at a specific current set point by the procedure outlined in Fig.\ \ref{fig:SI1}. We extract the magnetic field positions of the ESR signal maxima at each bias voltage from a magnetic field/bias voltage sweep and do a spline fit of the bias voltage vs.\ magnetic field points as shown in Fig.\ \ref{fig:SI1}\panel{a}. In practice, bias voltages smaller than $\pm$20\,mV do not show a clear ESR signal, which we attribute to too low currents close to zero bias voltage. To bridge this gap, we interpolate the missing data points with a spline fit. We then use the spline fits at four different microwave frequencies and perform a linear fit at each bias voltage. The spline fit data for each microwave frequency (i.e.\ Zeeman energy) is plotted in Fig.\ \ref{fig:SI1}\panel{b} along with a linear fit at two bias voltages ($\pm$150\,mV). Each linear fit yields the $g$-factor (slope) and the tip field $B_\text{tip}$ (negative $x$-axis intercept) found at that bias voltage. From there, we can plot the dependency of the $g$-factor and tip field $B_\text{tip}$ with respect to the bias voltage that we have presented in Fig.\ 2 of the main text.

\begin{figure}
\centerline{\includegraphics[width=\columnwidth]{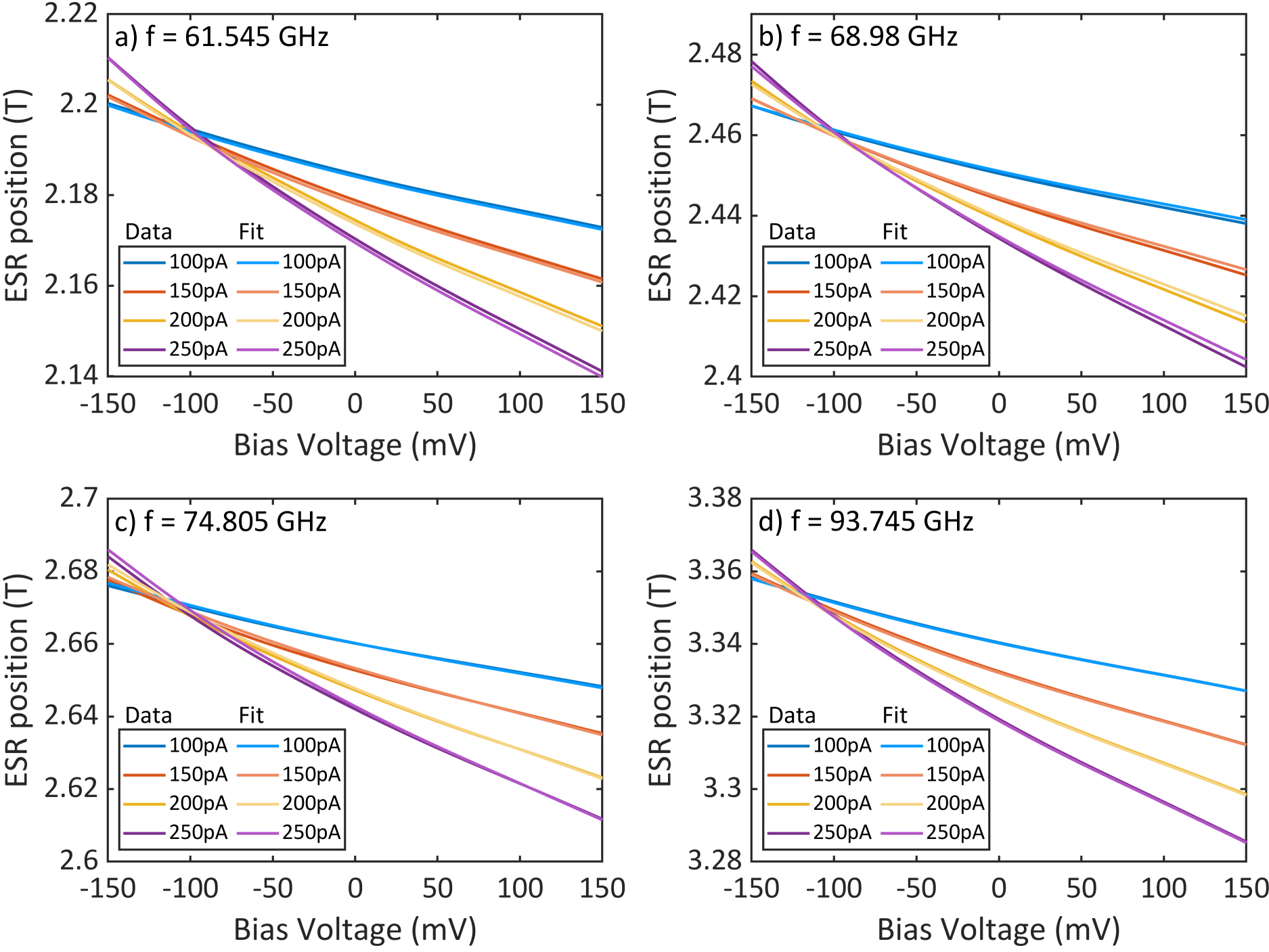}}
\caption{Comparison of the extracted ESR peak positions with the peak positions calculated from the fitted $g$-factors and tip fields $B_\text{tip}$. The different panels show the different microwave frequencies. We find generally good agreement for all frequencies and current setpoints.}
\label{fig:modelpeaks}
\end{figure}

\begin{figure}
\centerline{\includegraphics[width=\columnwidth]{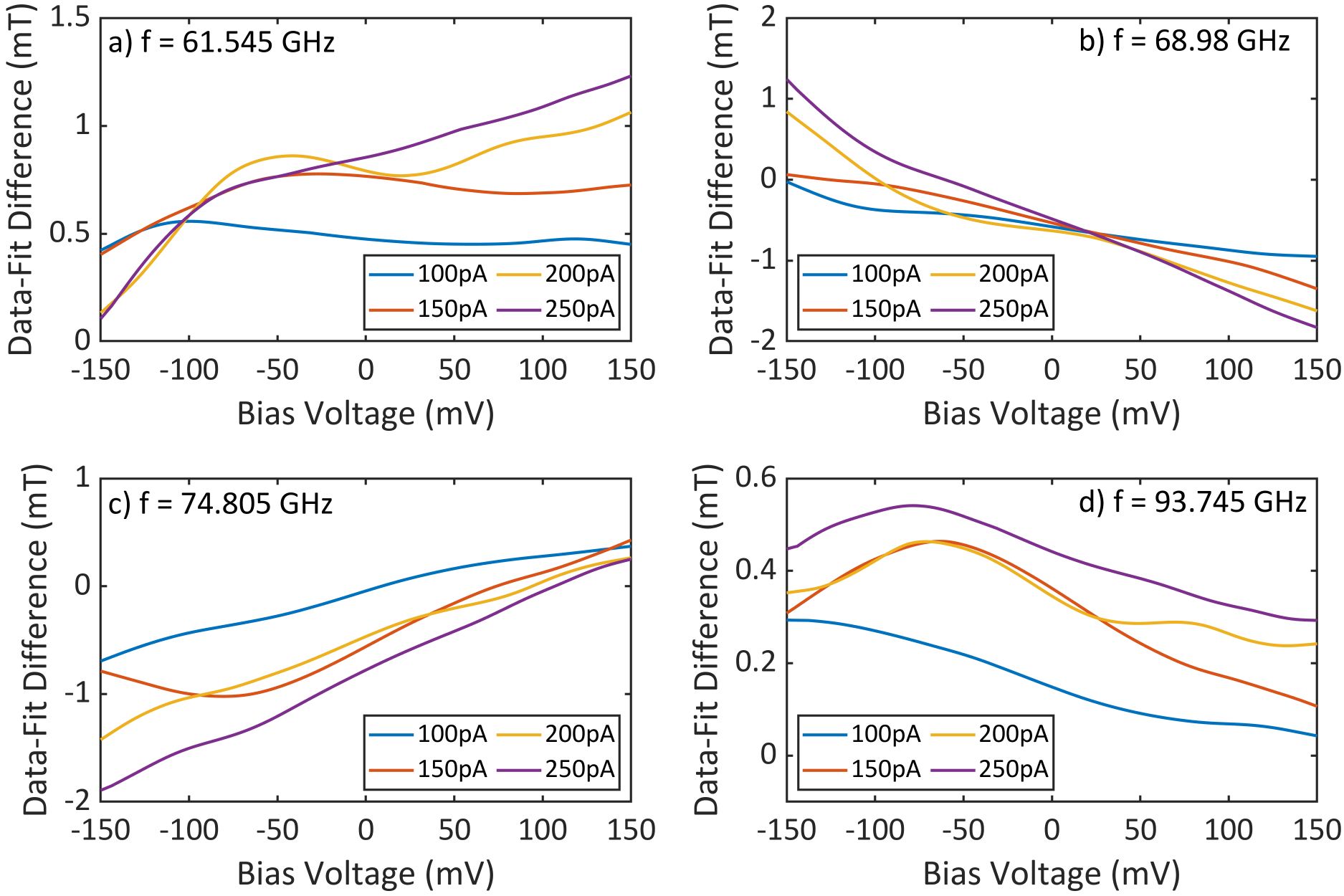}}
\caption{Differences between the extracted and the calculated ESR peak positions shown in Fig.\ \ref{fig:modelpeaks}. The deviations are always less than 2\,mT indicating good overall agreement.}
\label{fig:fitdiff}
\end{figure}

To demonstrate the overall consistency of this multidimensional fit, we plot the extracted ESR peak positions along with the ESR resonance positions calculated from the fitted values, which are shown in Fig.\ \ref{fig:modelpeaks} for all microwave frequencies and current setpoints. We see an overall good agreement and a continuous evolution. To illustrate the agreement quantitatively, we calculate the difference between the experimental data and the modeled resonances, which is shown in Fig.\ \ref{fig:fitdiff} for the corresponding data in Fig.\ \ref{fig:modelpeaks}. We see that the deviations are generally small and never exceeding 2\,mT. Therefore, we conclude that we have an overall consistent model.

\section{Tip Approach}

\begin{figure}
\centerline{\includegraphics[width=\columnwidth]{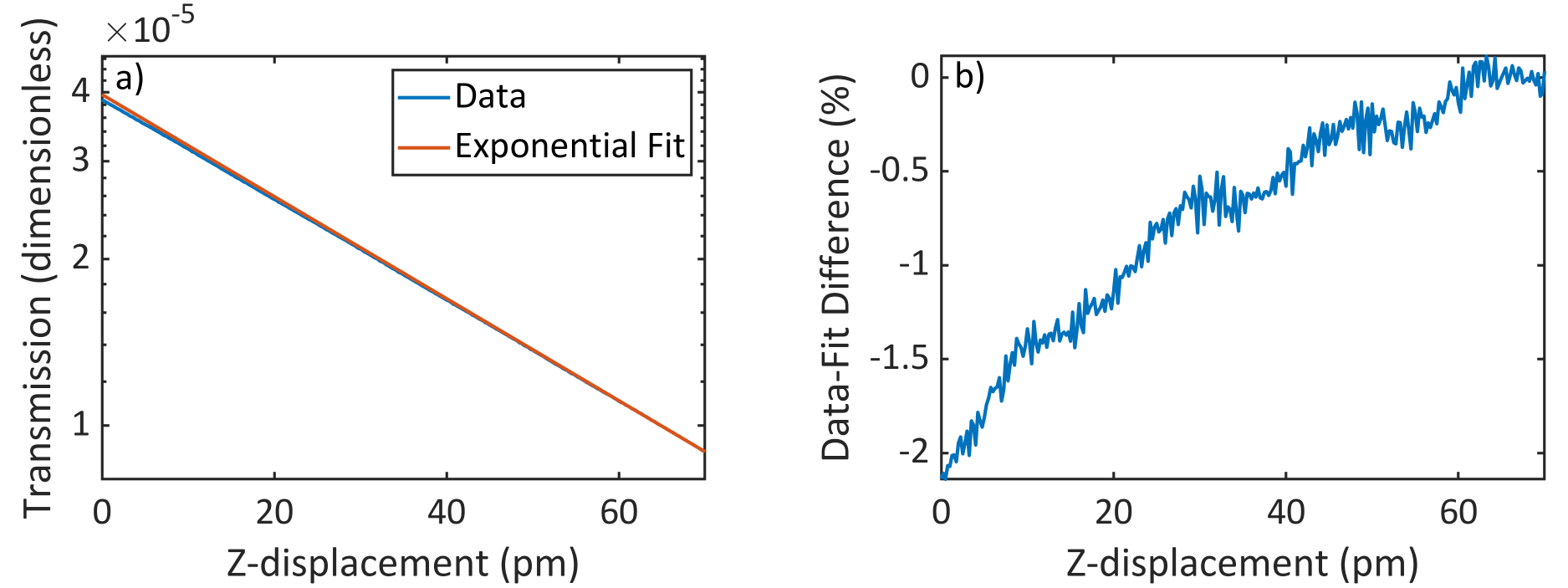}}
\caption{\panelcaption{a} Junction transmission as a function of $z$-displacement (tip-sample distance). The exponential fit is done in the low transmission regime revealing the sub-exponential evolution of the data. \panelcaption{b} Difference between data and fit normalized to the fit showing the sub-exponential evolution of the tip-approach.}
\label{fig:iz}
\end{figure}

The coupling of the TiH molecule to the substrate can be inferred from the evolution of the tunnel junction transmission as a function of the tip sample distance. A similar situation has been analyzed previously in a different context \cite{si_huang_quantum_2020}. Assuming that the TiH molecule is coupled to the substrate by the molecule-substrate coupling $\Gamma_\text{s}$ and to the tip by the molecule-tip coupling $\Gamma_\text{t}$, the junction transmission $\tau$ can be written as \cite{si_huang_quantum_2020,si_cuevas_molecular_2010}
\begin{equation}
    \tau = \frac{4\Gamma_\text{s}\Gamma_\text{t}}{(\Gamma_\text{s}+\Gamma_\text{t})^2}\overset{\Gamma_\text{t}\ll\Gamma_\text{s}}{=}\frac{4\Gamma_\text{t}}{\Gamma_\text{s}}.
\end{equation}
The transmission $\tau$ describes the junction conductance in units of the quantum of conductance $G_0=2e^2/h$, where $e$ is the electron charge and $h$ is Planck's constant. Since our junction is in the tunneling limit, i.e.\ $\Gamma_\text{t}\ll\Gamma_\text{s}$, we can easily see that a change in the molecule-substrate coupling $\Gamma_\text{s}$ has a direct impact on the evolution of the junction transmission. We can reasonably assume that in the tunneling regime, the molecule-tip coupling $\Gamma_\text{t}$ increases exponentially with decreasing tip-sample distance. If the molecule-substrate coupling $\Gamma_\text{s}$ increases/decreases as the tip-sample distance decreases, the junction transmission $\tau$ will evolve less/more than exponentially, respectively. The tip approach for the tunnel junction measured in the main text in Fig.\ 2 is shown in Fig.\ \ref{fig:iz}\panel{a}. The blue line represents the data, while the red line represents an exponential fit to the data points at $z$-positions $>60\,$pm. A small but clear subexponential deviation of the data can be seen. The relative difference between data and fit is also plotted in Fig.\ \ref{fig:fitdiff}\panel{b} indicating that the junction transmission evolves below the fitted exponential evolution. From this behavior, we conclude that the molecule-substrate coupling $\Gamma_\text{s}$ increases as the tip approaches the molecule. Therefore, it is likely that the molecule is pushed towards the surface in this approach range. This provides an overall consistent picture of an increasing $g$-factor as the molecule-substrate distance decreases \cite{si_steinbrecher_quantifying_2021} and explains the evolution of the $g$-factor at zero bias voltage for decreasing tip-sample distance. 

\section{Modelling Coupled Spins}

\begin{table}
\begin{tabular}{ |p{1cm}||p{1.4cm}|p{1.4cm}|p{1.9cm}|p{1.9cm}|  }
\hline
 \multicolumn{1}{|c||}{}&\multicolumn{2}{|c|}{Avoided Crossing}&\multicolumn{2}{|c|}{Large $J$} \\ [1ex]
 \hline
 $I_\text{sp}$ & 400\,pA & 400\,pA & 1\,nA & 1\,nA \\    
 $U_\text{sp}$ & 100\,mV & 100\,mV & 150\,mV & 150\,mV \\ 
 \hline
 $U_\text{bias}$ & 0\,mV & 200\,mV & 0\,mV & 200\,mV\\
 \hline
 $J$   & 0.669\,GHz   & 0.669\,GHz &   61.11645\,GHz & 61.11645\,GHz\\
 $D$   & 13.3\,MHz   & 13.3\,MHz &   50\,MHz & 50\,MHz\\
 $g_1$   & 1.92    & 1.925 &   2.0703 & 2.103\\
 $g_2$   & 1.975    & 1.973 &   1.87 & 1.913\\
 $B_\text{tip}$   & 0\,mT    & 14\,mT &   20\,mT & 36\,mT\\
 \hline
\end{tabular}
\caption{Fit parameters for the two dimers presented in the main text. The bias voltage values in the table are the extremal values at the edge of the interval. The parameters for the bias voltage values in between are linearly interpolated. The corresponding $g$-factors and tip fields differ between the dimers because they were measured with different tips and at different current and voltage set points $I_\text{sp}$ and $U_\text{sp}$.}
\label{tab:params}
\end{table}

The models presented in Fig.\ 3\panel{b} and \panel{c} of the main text are based on a coupled spin Hamiltonian found in literature \cite{si_veldman_free_2021,si_bae_enhanced_2018,si_yang_engineering_2017}:
\begin{equation}
\begin{split}
H = - \mu_\text{B} (B_\text{ext} + B_\text{tip}) g_1 \hat{S}^z_1 - \mu_\text{B} B_\text{ext}  g_2 \hat{S}^z_2 \\ + J \hat{{S}}_1 \cdot \hat{{S}}_2 + D(3\hat{S}^z_1\hat{S}^z_2 -\hat{{S}}_1 \cdot \hat{{S}}_2).
\end{split}
\label{eq:SpinHam}
\end{equation}
\noindent This Hamiltonian works on the spin operators of the coupled spins, where $g_1$ and $g_2$ are the $g$-factors of the TiH molecule beneath the tip and beside the tip, respectively, $B_\text{tip}$ is the tip field that is only considered to affect the TiH molecule beneath the tip, $B_\text{ext}$ is the external magnetic field, $J$ is the Heisenberg interaction energy between the two spins, and $D$ is the dipole interaction between the two spins. For the dipole interaction, we estimate $D = 13.3$\,MHz for the more distant dimer and $D=50\,$MHz for the closer dimer, which is a significantly smaller contribution than the other interactions.

\begin{figure}
\centerline{\includegraphics[width=0.7\columnwidth]{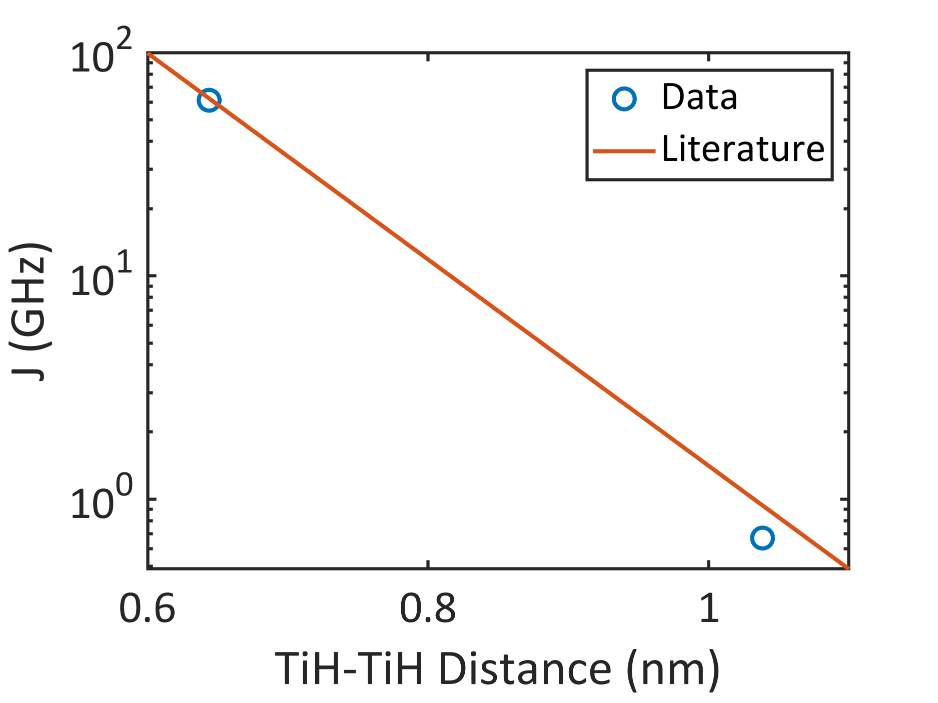}}
\caption{Comparison of the fit parameters for the exchange coupling $J$ in the two dimers with the exponential dependence reported in literature \cite{si_bae_enhanced_2018}.}
\label{fig:jcoupling}
\end{figure}

Modelling of the experimentally observed transitions is done by considering the energy difference between two eigenvalues of the spin Hamiltonian. Modeling the TiH molecules as \spinhalf\ systems, the spin Hamiltonian in Eq.\,\eqref{eq:SpinHam} can be diagonalized to analytically find the four eigenstates, three triplet states ($\vert \text{T\textsubscript{+}}\rangle$, $\vert \text{T\textsubscript{0}}\rangle$ and $\vert \text{T\textsubscript{-}}\rangle$) and one singlet state ($\vert \text{S}\rangle$). In the case of the avoided crossing, we observe four transitions with energies: $|E_{T_0}-E_{T_-}|$, $|E_{T_0}-E_{T_+}|$, $|E_{S}-E_{T_-}|$, and $|E_{S}-E_{T_+}|$ \cite{si_zhang2022electron,si_veldman_free_2021}. We then equate these energy differences to the energy $hf$ of the microwave radiation, which leads to the following set of equations:
\begin{multline}
hf = |E_{T_O}-E_{T_+}| = \frac{1}{2}[J+2D+\mu_\text{B}(g_1(B_\text{ext}+B_\text{tip})+g_2B_\text{ext}) \\ 
-\sqrt{(J-D)^2+(\mu_\text{B}(g_1(B_\text{ext}+B_\text{tip})+g_2B_\text{ext}))^2}],
\end{multline}\label{eq:SpinHam2}
\begin{multline}
hf = |E_{T_O}-E_{T_-}| = \frac{1}{2}[J+2D-\mu_\text{B}(g_1(B_\text{ext}+B_\text{tip})+g_2B_\text{ext}) \\ 
-\sqrt{(J-D)^2+(\mu_\text{B}(g_1(B_\text{ext}+B_\text{tip})+g_2B_\text{ext}))^2}],
\end{multline}\label{eq:SpinHam3}
\begin{multline}
hf = |E_{S}-E_{T_+}| = \frac{1}{2}[J+2D+\mu_\text{B}(g_1(B_\text{ext}+B_\text{tip})+g_2B_\text{ext}) \\ 
+\sqrt{(J-D)^2+(\mu_\text{B}(g_1(B_\text{ext}+B_\text{tip})+g_2B_\text{ext}))^2}],
\end{multline}\label{eq:SpinHam4}
\begin{multline}
hf = |E_{S}-E_{T_-}| = \frac{1}{2}[J+2D-\mu_\text{B}(g_1(B_\text{ext}+B_\text{tip})+g_2B_\text{ext}) \\ 
+\sqrt{(J-D)^2+(\mu_\text{B}(g_1(B_\text{ext}+B_\text{tip})+g_2B_\text{ext}))^2}].
\end{multline}\label{eq:SpinHam5}
\noindent Using this set of equations we can solve for four different external magnetic fields $B_\text{ext}$ ($B_{T_0T_+}$, $B_{T_0T_+}$, $B_{ST_+}$ and $B_{ST_-}$) numerically using input values for $J$, $D$, $g_1$, $g_2$ and $B_\text{tip}$. The microwave frequency $f$ is a known input quantity. The resulting external magnetic field values $B_\text{ext}$ represent the positions of the ESR peaks on the external magnetic field axis for the given transitions. In the case of the dimer with the stronger interaction energy, we model the positions of the ESR peaks by considering the following transition energies: $|E_{T_0}-E_{T_-}|$, $|E_{T_0}-E_{T_+}|$ and $|E_{S}-E_{T_0}|$ \cite{si_bae_enhanced_2018}.

To incorporate the effect of the bias voltage in the modelling we assume a linearly dependence of $g_1$, $g_2$, and $B_\text{tip}$ on the bias voltage in the range from 0\,mV to $200$\,mV. This is based on the results presented in Fig.\ 2\panel{c} and \panel{d} of the main text. We find that to get accurate results, $g_2$ also has to shift with the bias voltage, which implies that the electric field of the tip still affects the TiH molecule next to the tip apex. This is to be expected as the tip and sample can be approximated as a plate capacitor close to the tip apex. Furthermore, we assume that $J$ and $D$ are not affected by the bias voltage. We choose the values for $J$ according to the exponential distance dependence between the molecules in the dimer that has been established previously \cite{si_bae_enhanced_2018,si_yang_probing_2021,si_yang_engineering_2017}. The comparison is shown in Fig.\ \ref{fig:jcoupling}, where the red line is given by $J=J_0\exp\left(-(r-r_0)/d\right)$ with $r_0=0.72\,$nm, $d = 94\,$pm, and $J_0 = 27.7\,$GHz \cite{si_yang_probing_2021}. We then input a constant $J$ and $D$ into our set of equations and solve for the external magnetic field values $B_\text{ext}$ constituting the positions of the ESR peaks as described above. For each set of $g_1$, $g_2$ and $B_\text{tip}$, we solve for $B_{T_0T_+}$, $B_{T_0T_+}$, $B_{ST_+}$ and $B_{ST_0}$ for the case of the dimer with the avoided crossing, and $B_{T_0T_+}$, $B_{T_0T_+}$ and $B_{ST_0}$ for the case of the dimer with a larger interaction energy. Finally, we superimpose the calculated ESR transitions over the data to find the parameters with the best fit. The fit parameters $g_1$, $g_2$, $B_\text{tip}$, $D$ and $J$ for the two different dimers are presented in Table \ref{tab:params}.

To plot the modelled eigenergies in Fig.\ 3\panel{b} and \panel{c} of the main text, we simply input our estimated values for $J$ and linearly changing $g_1$, $g_2$ and $B_\text{tip}$ into the diagonalized eigenergies of Eq.\,\eqref{eq:SpinHam}. Here, the $x$-axis in Fig.\ 3\panel{b} and \panel{c} is an ``effective'' bias voltage that we model with linearly shifting values for $g_1$, $g_2$ and $B_\text{tip}$, but for a constant external magnetic field $B_\text{ext}$. Therefore, the evolution of the energy levels in Fig.\ 3\panel{b} and \panel{c} and the experimental data in Fig.\ 3\panel{a} and \panel{d} are not directly comparable. 

\section{Measurements on TiH\textsubscript{O}}

\begin{figure}[h]
\centerline{\includegraphics[width=\columnwidth]{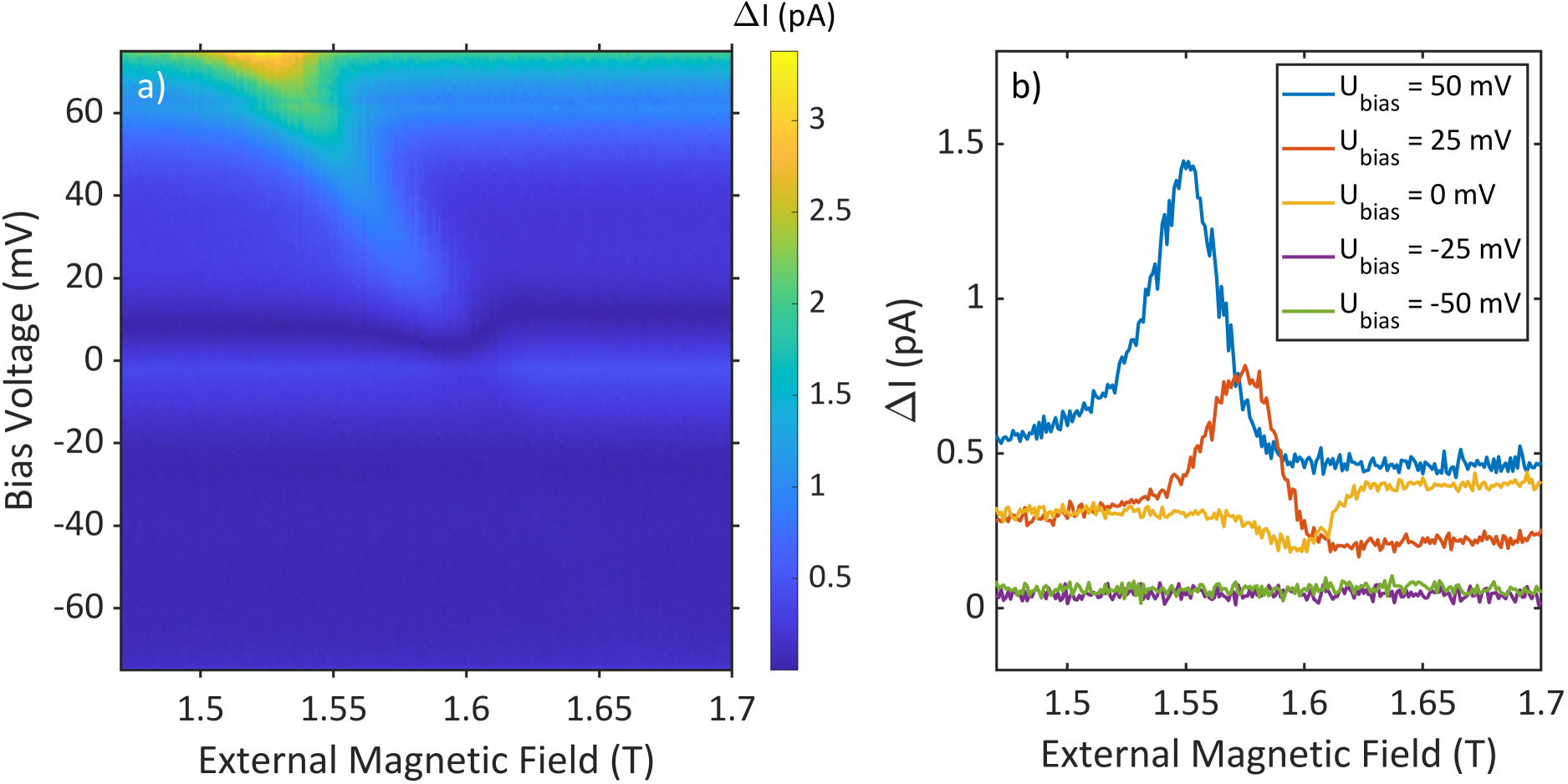}}
\caption{\panelcaption{a} Magnetic field/bias voltage sweep measured on a TiH\textsubscript{O} molecule ($U_\text{sp} = 100$\,mV, $I_\text{sp} = 75$\,pA, $f_\text{rf} = 19$\,GHz, $U_\text{rf} = 20$\,mV). \panelcaption{b} ESR sweeps measured on TiH\textsubscript{O} at different bias voltages. The evolution of the ESR peak shows a similar voltage dependence as for TiH\textsubscript{OO}, but the spin-electric coupling is stronger.}
\label{fig:SI2}
\end{figure}

Magnetic field/bias voltage sweeps were also performed on on-site TiH molecules (TiH\textsubscript{O}). Fig.\ \ref{fig:SI2}\panel{a} shows such a sweep where the ESR signal can be clearly seen. We see a linear shift of the ESR peak at positive bias voltages and no signal at negative bias voltages. We found that increasing the set point current of the magnetic field/bias voltage sweeps on TiH\textsubscript{O} increased the linear shift of the ESR signal with respect to the bias voltage, which is consistent with our observations on TiH\textsubscript{OO} molecules. We found that the shift of the TiH\textsubscript{O} is much stronger than on the TiH\textsubscript{OO} for similar setpoint currents. We attribute this to the tip being closer to the sample when measuring on TiH\textsubscript{O} than when measuring on TiH\textsubscript{OO}. This is due to the smaller local density of states on the TiH\textsubscript{O} molecule, which leads to the tip sample distance being smaller on the TiH\textsubscript{O} than on the TiH\textsubscript{OO} for comparable set points. This is supported by the different appearance of the TiH\textsubscript{O} compared to the TiH\textsubscript{OO} as the TiH\textsubscript{OO} molecules appear brighter than the TiH\textsubscript{O} molecules (cf.\ Fig.\ 1\panel{a} of the main text).

\section{Spin-Electric Coupling for Different Tips and TiH\textsubscript{OO} Molecules}

\begin{figure}[t]
\centerline{\includegraphics[width=\columnwidth]{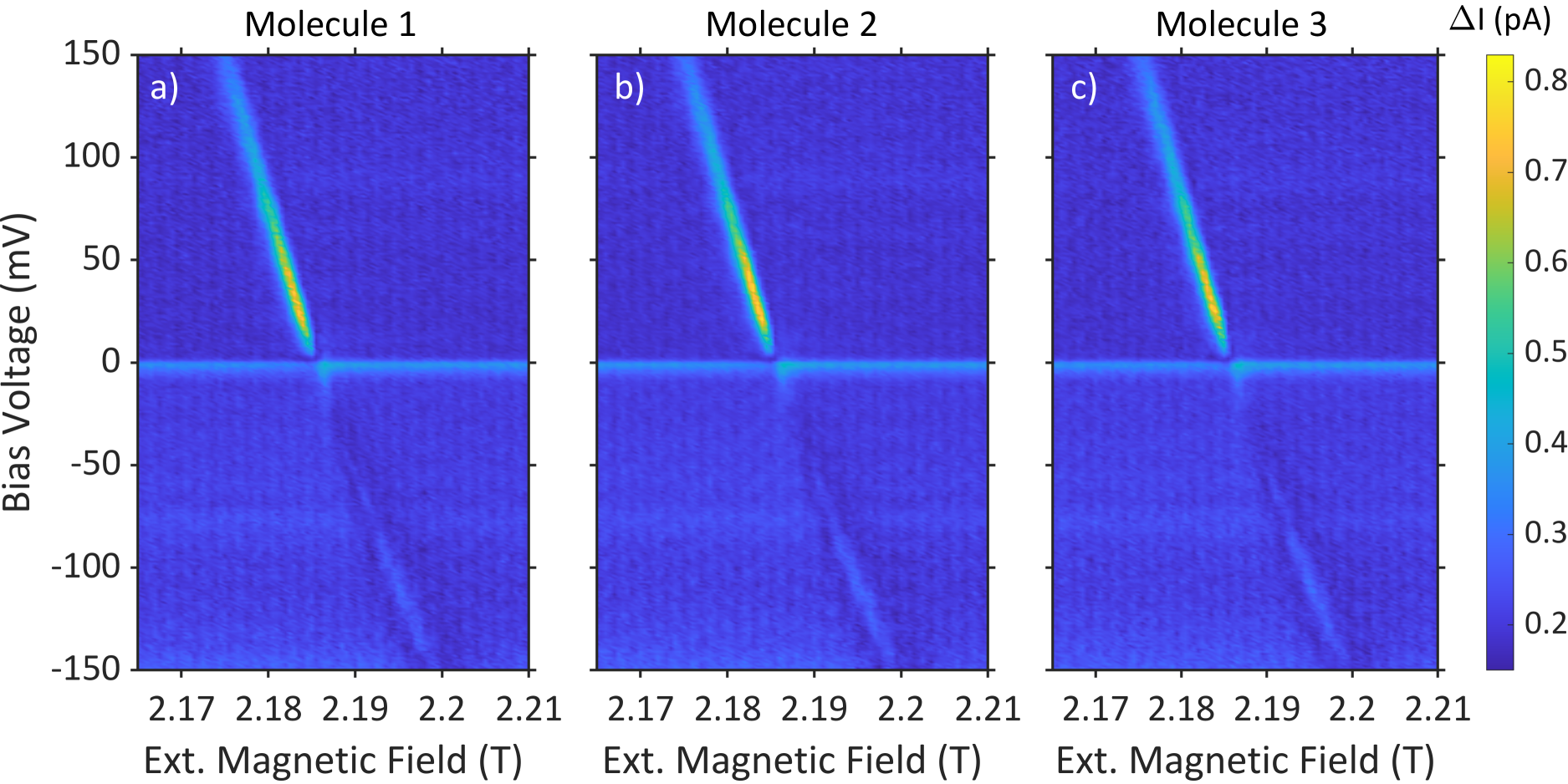}}
\caption{Magnetic field/bias voltage sweeps measured on three different TiH\textsubscript{OO} molecules ($U_\text{sp} = 100$\,mV, $I_\text{sp} = 100$\,pA, $f_\text{rf} = 61.545$\,GHz, $U_\text{rf} = 20$\,mV). All three TiH\textsubscript{OO} molecules show the same behavior.}
\label{fig:SI3}
\end{figure}

As a consistency check we performed magnetic field/bias voltage sweeps on various TiH\textsubscript{OO} molecules found on the sample with the same tip. Fig.\ \ref{fig:SI3} shows three such sweeps on three different TiH\textsubscript{OO} molecules from which we conclude that the measurements are consistent and reproducible. 

Furthermore, the data presented in the main text was measured using two different ESR-functionalized tips. The first tip was used for the measurements shown in Fig.\ 1, Fig.\ 2 and Fig.\ 3\panel{a}. The second tip was used in the avoided crossing measurements shown in Fig.\ 3\panel{d} and Fig.\ 4. In addition, over the course of this study we have observed spin-electric coupling in the ESR signal for five tips. Lastly, during our experiments we never encountered an ESR tip nor a TiH molecule that did not show spin-electric coupling.

\end{document}